# Time-Shift in the OPERA set-up: proof against superluminal neutrinos without the need of knowing the CERN-LNGS distance and Reminiscences on the origin of the Gran Sasso Lab, of the 3rd neutrino and of the "Teramo Anomaly"

Antonino Zichichi
*INFN and University of Bologna, Italy*
*CERN, Geneva, Switzerland*

# Time-Shift in the OPERA set-up: proof against Superluminal neutrinos without the need of knowing the CERN-LNGS distance and Reminiscences on the origin of the Gran Sasso Lab, of the 3$^{rd}$ neutrino and of the "Teramo Anomaly"

A. Zichichi

## Abstract

The LVD time stability allows to establish a time-shift in the OPERA experiment, thus providing the first proof against Superluminal neutrinos, using the horizontal muons of the "Teramo Anomaly". This proof is particularly interesting since does not need the knowledge of the distance between the place where the neutrinos are produced (CERN) and the place where they are detected (LNGS). Since the Superluminal neutrinos generated in the physics community a vivid interest in good and bad behaviour in physics research, the author thought it was appropriate to recall the origin of the Gran Sasso Lab, of the 3$^{rd}$ neutrino, of the horizontal muons due to the "Teramo Anomaly" and of the oscillation between leptonic flavours, when the CERN–Gran Sasso neutrino beam was included in the project for the most powerful underground Laboratory in the world.

## Table of Contents

# Time-Shift in the OPERA set-up: proof against superluminal neutrinos without the need of knowing the CERN-LNGS distance and Reminiscences on the origin of the Gran Sasso Lab, of the 3rd neutrino and of the "Teramo Anomaly"

A. Zichichi

## 1 INTRODUCTION

The purpose of this work is to report the measurement of a time-shift in the OPERA set-up in a totally independent way from TOF measurements of CNGS neutrino events and without the need of knowing the distance between the two laboratories, CERN and LNGS, where the neutrinos are produced and detected, respectively. This can be done thanks to the existence of the "Teramo anomaly". The LVD and OPERA experiments are both installed in the same laboratory: LNGS. Indeed, the OPERA-LVD direction lies along the so-called "Teramo anomaly", a region in the Gran Sasso massif where LVD has established, many years ago, the existence of an anomaly in the mountain structure, which exhibits a low $m.w.e.$ thickness for horizontal directions.

The "abundant" high-energy horizontal muons (nearly 100 per year) going through LVD and OPERA exist because of this anomaly in the mountain orography. It is these muons which allow to have a time-correlation, between different set-ups.

In this lecture after the "Teramo anomaly", discussed in Chapter 2, I will recall the origin of the 3rd neutrino (Chapter 3) since the purpose of OPERA was (and is) to measure the oscillations between the neutrinos of the second and third family, the last one being now called $\nu_\tau$, but originally designated as $\nu_{HL}$ where HL stands for Heavy Lepton. In Chapter 4 and Chapter 5 few words will be devoted to the two set-ups LVD and OPERA, respectively. The crucial measurement of the time-shift between LVD and OPERA will be treated in Chapter 6. The Conclusions in Chapter 7. Acknowledgment in Chapter 8.

## 2 The discovery of the "Teramo Anomaly" (an example of how to avoid big mistakes) and the problem of experimental versus theoretical papers

How did we discover this "anomaly" is an example of how to avoid big-mistakes [1]. The problem was to **search** for Galactic Center Supernovae (SN). **Why**? Because from Keplero–Galilei last **Supernova (SN)** there are four centuries. The rate of SN per Galaxy per century is **3**. The number of missing SN is twelve.

How can this be explained? **Answer**: the **light** emitted by **SN** is what we observe. But the centre of our Galaxy is with enormous light emission. **Conclusion**: the centre of our Galaxy has to be observed not thanks to the light emitted by a SN but by **neutrino-emission-effects**.

Notice that all Supernovae known so far have been detected thanks to light but they are all in a galactic-space where the number Stars is $\simeq 10\%$ of the total number. Therefore the number of missing SN in four centuries is not 12 but $\simeq 1,2$. The number of missing SN would be 12 if the telescopes for neutrinos from SN were active during the past four centuries.

**But** it is only since 1992 that we have **the LVD telescope** being active for the observation of SN-neutrino. Since the expected number is one Supernova every 30 years, the number of years of active LVD being only 20, the effective number of missing SN when the observation includes the neutrino is $\simeq 0,7$.

The number of SN expected to be observed via light mission during four centuries (1,2) and the number of SN expected to be observed by ν telescopes during the past 20 years (0,7) are in purely accidental agreement. If we are lucky we should observe with LVD a SN in the next years to come. This is why when we started with our LVD telescope we focused our attention to look toward the centre of our Galaxy: not only for SN neutrinos but also as a source of unexpected events giving rise to any sort of anomaly in our detectors. Looking at the centre of our Galaxy we discovered the "Teramo anomaly", i.e. high-energy muon flux $\simeq 100$ per year.

This anomaly could have produced a paper where the effect was attributed to what happens in the centre of our Galaxy. But we are not theorist whose model can be wrong and nothing happens. In experimental physics it is forbidden to make mistakes. Being member of the Blackett School, instead of being excited by a fast conclusion, we decided to go on in our search. This is how we discovered that the centre of our Galaxy had nothing to do with the intense flux of muons (the Teramo anomaly). This flux was due to an unexpected anomaly in the mountain density. This "beam" of muons could be of great value for the Gran Sasso experiments, as proved by our meeting here.

A sketch of the LNGS map with the position of LVD and OPERA experiments is shown in Figure 1.

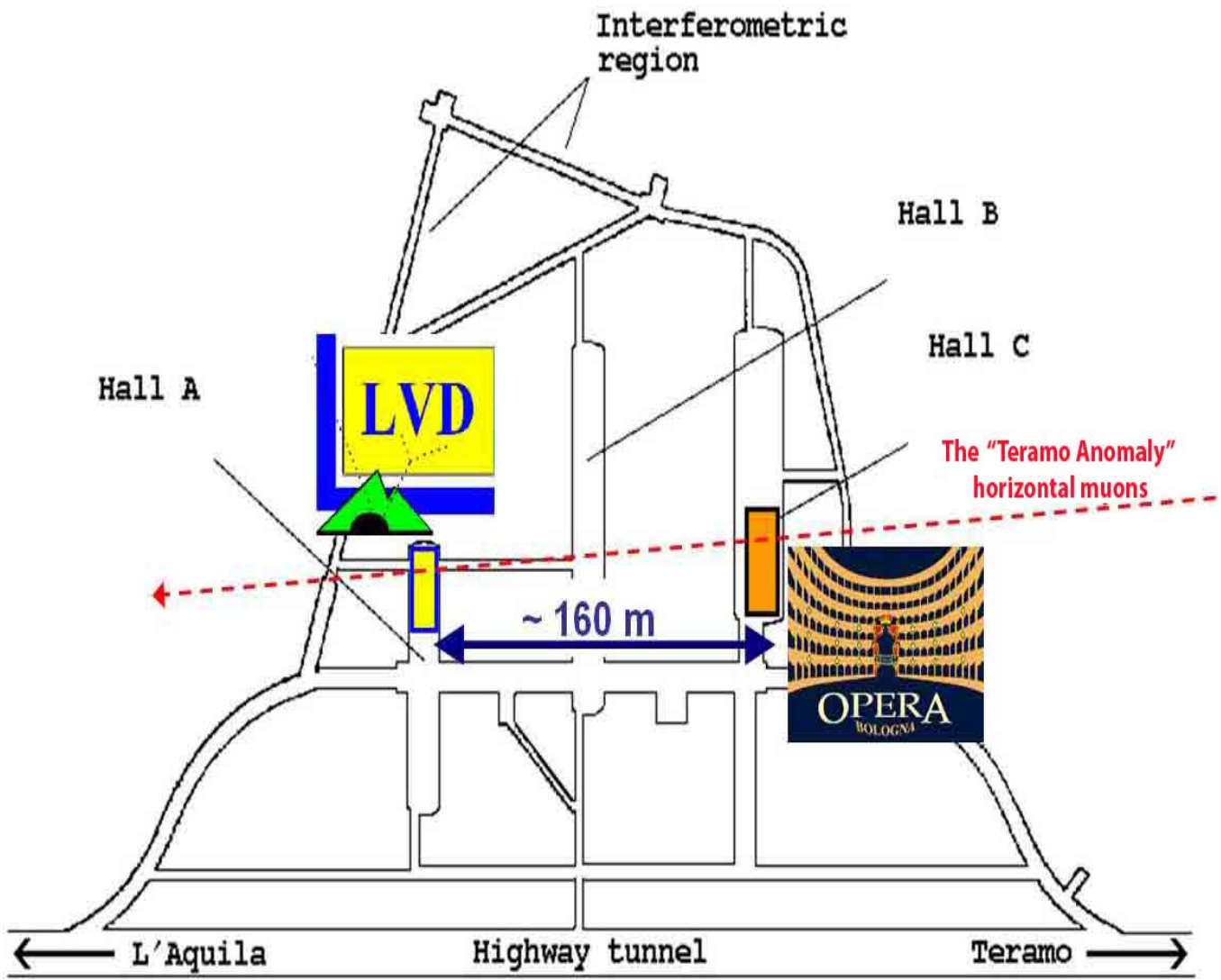

*Fig. 1: Sketch of the LNGS map with the position of the LVD and OPERA experiments.*

Indeed, the OPERA-LVD direction lies along the anomaly in the mountain profile observed in 1997 [1] when searching for neutrino events from the centre of the Galaxy.

The anomaly is due to the non-uniform rock structure in the horizontal direction towards the city of Teramo, thus called "Teramo anomaly".

This is due to a large decrease in $m.w.e.$ (meter, water, equivalent) of the mountain rock structure, as indicated in Figure 2 by the red circle.

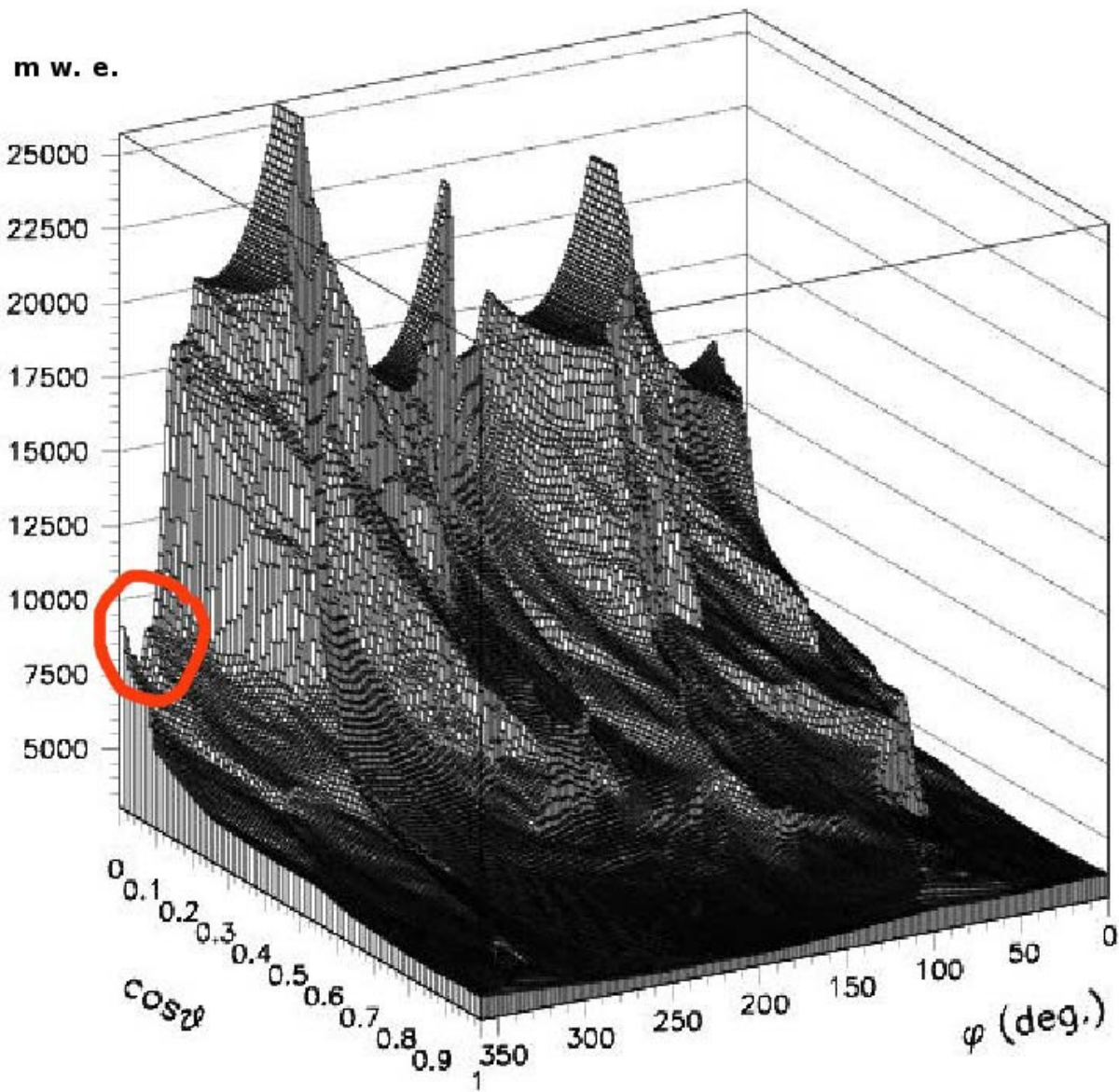

*Fig. 2: Map of the slant depth of the Gran Sasso mountain (in m.w.e.) as a function of the arrival direction: θ and ϕ are respectively the zenith and azimuth angle. The red circle indicates the direction of the "Teramo valley", where the mountain profile exhibits an "anomaly" in the m.w.e. for horizontal directions.*

Let me say few words to explain my previous Statement of being member of the Blackett group. When I started my research activity I had the privilege of being a pupil of Professor Patrick M.S. Blackett whose famous Statement was: *«We experimentalists are not like theorists: the originality of an idea is not for being printed in a paper, but for being shown in the implementation of an original experiment»* (London 1962).

This Statement (Figure 3) is at the entrance of the Blackett Institute of the EMFCSC in Erice.

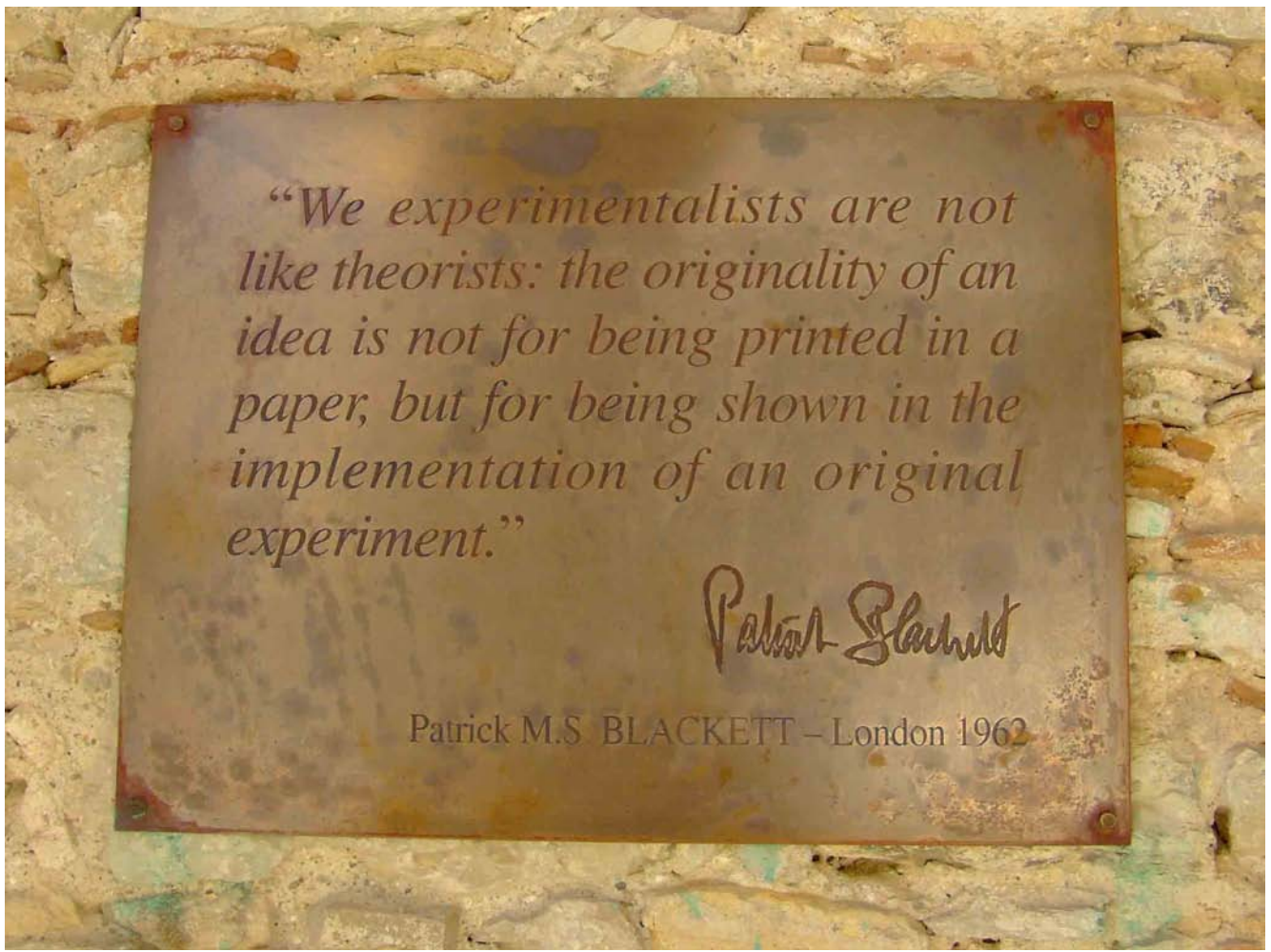

*Fig. 3.*

And here comes another crucial point which distinguishes experimental physics papers from theoretical physics papers.

Many years ago there was a Seminar at CERN devoted to this problem [2]. During the discussion I quoted the view point of Feynman about punishing the authors of models which had no relation with physics reality and Martinus Veltman expressed his support. As every body knows this view-point is the basis of the famous statement by Wolfgang Pauli «*I don't mind that you think slowly but I do mind that you are publishing faster than you think*». Unfortunately the status of experimental versus theoretical paper is as given below:

a) **Theorist** ⇒ if the **model** is not corroborated by an experimental result: nothing happens. The theoretical physicist does not **lose his reputation**.

b) **Experimentalist** ⇒ if the experimental result is not corroborated by other experimental results, the experimental physicist **loses his reputation**.

This is the reason why we, experimental physicists, must be extremely careful in letting people knowing – even confidentially – if we find what could be a very interesting result. In fact the more the result is interesting, the more we must be careful. Also because you can never be sure that, talking to a colleague, he is going to keep the secret for himself. The lesson from the past is very instructive. Let me quote the case when the proton was proposed to be the antielectron by Dirac; Kapitza was responsible for this, as I report in the Erice folder (distributed to all participants every year since 1963). I want to skip this amusing example since it is known to all physicists coming to Erice, while very few people known the origin of the OPERA main physics task: to study the oscillations between the two neutrinos $\nu_\mu$ and $\nu_\tau$.

## 3 THE ORIGIN OF THE 3$^{rd}$ NEUTRINO AND SOME REMINISCENCES OF HOW THE GRAN SASSO LAB COULD BECOME REAL

What is now called $\tau$–lepton represents the origin of the 3$^{rd}$ family of fundamental fermions. The search for a lepton heavier than the muon started at CERN with the PAPLEP (P̲roton A̲ntiproton Annihilation into L̲epton P̲airs) ($p\bar{p}$) experiment in the early sixties, before the discovery in 1964 of PC-violation by J. Christenson, J.W. Cronin, Val L. Fitch, and R. Turlay [3] and the proposal in 1973 by Kobayashi and Maskawa [4] for the existence of a 3rd family of fundamental fermions.

Since the scientific motivation for OPERA is to detect the oscillation between the 2$^{nd}$ and the 3$^{rd}$ neutrino, it is appropriate to establish the origin of this physics.

In Figure 4 there is the cover page of the volume due to M.me C.S. Wu to whom I want to renew my gratitude.

In Figure 5 the first experimental set-up constructed in the early sixties at CERN for the search of the 3$^{rd}$ lepton, called by us at that time HL, is reported.

*Fig. 4.*

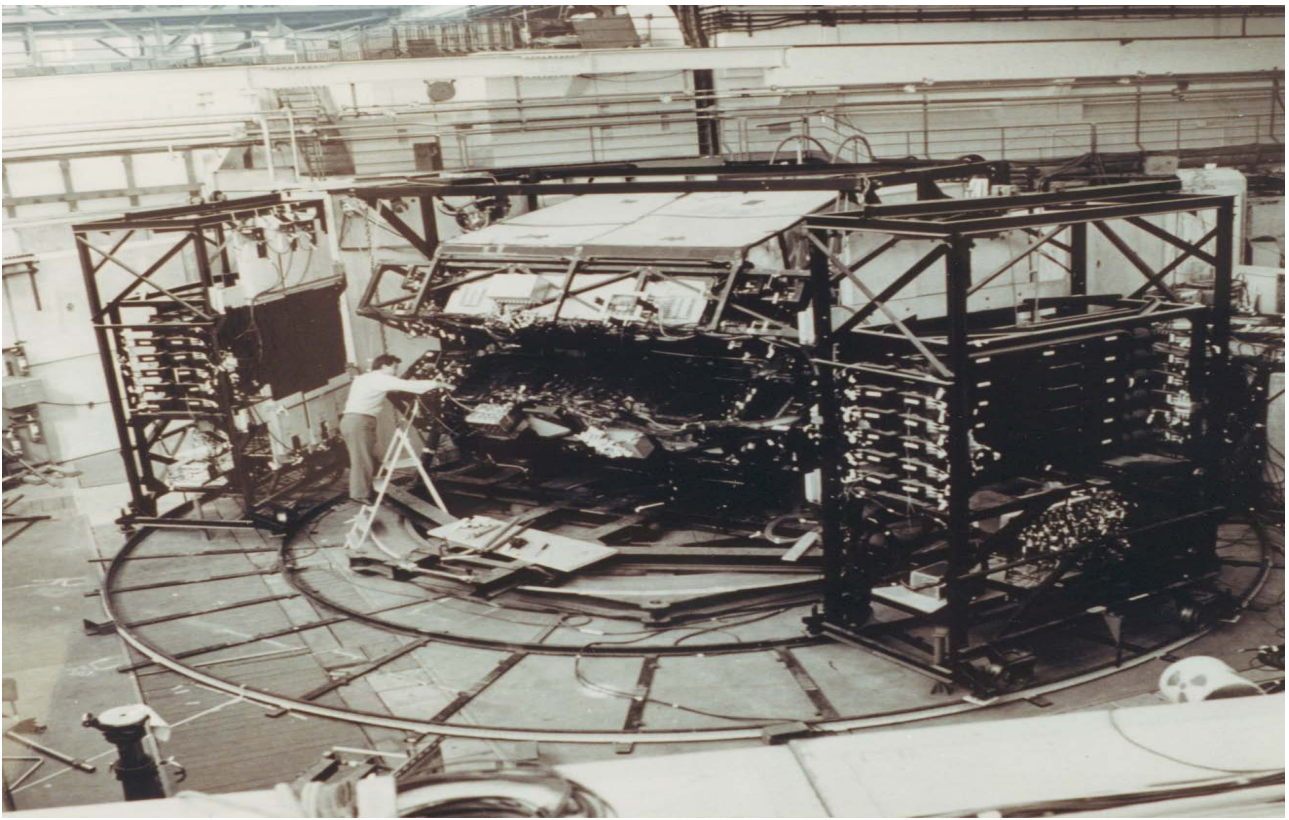

*Fig. 5: The PAPLEP experimental set-up at CERN.*

This set-up is the first example of the gigantic set-ups now needed to do physics as it is the OPERA detector. From PAPLEP to OPERA there are 40 decades; nevertheless the search for the 3$^{rd}$ leptonic family started at CERN 40 decades ago using as production process ($p\bar{p}$) annihilation.

No-one knew the existence of the very large ElectroMagnetic Form Factor (EMFF) of the nucleon in the time-like region when PAPLEP was proposed. This large EMFF gave as result a depression factor more than 500 in the ($p\bar{p}$) annihilation cross-section for the productions of pairs of the third lepton (HL) ($\overline{HL}$). This is why the search moved to Frascati where the production process was ($e^+e^-$).

The search for the existence of a 3$^{rd}$ lepton at CERN was not a sporadic event. The possible existence of a lepton heavier than the muon had its roots in the research activity of this great new Laboratory in Europe, where attention was given to the validity of QED when particles 200 times heavier than the electrons were involved.

This is why the high precision determination of the muon anomalous magnetic moment of $(g–2)_\mu$ was performed. And the problem of establishing the universality of the weak forces was studied, with the first high precision measurement of the weak coupling, through the muon decay lifetime: $\tau_\mu$.

The reason for the abundance of muons was the incredible mass difference between the π–meson and its decay partner, the μ–lepton. The problem was that if this incredible coincidence did not repeat in the GeV–range. In this case a third lepton much heavier than the muon, HL, could be there but very rare since the sequence π → μ was unlikely to be there for the pair of heavy meson → heavy lepton.

A schematic view of the present status of the electroweak forces where the origin of the third family of leptons is correctly represented is in the bottom-left of Figure 6.

In this Figure the physics roots are dated 1960. These roots gave rise to the Gran Sasso Project (1979) where the study of oscillations between $\nu_\mu$ and $\nu_{HL}$ was indeed included, as testified by Pontecorvo (see later).

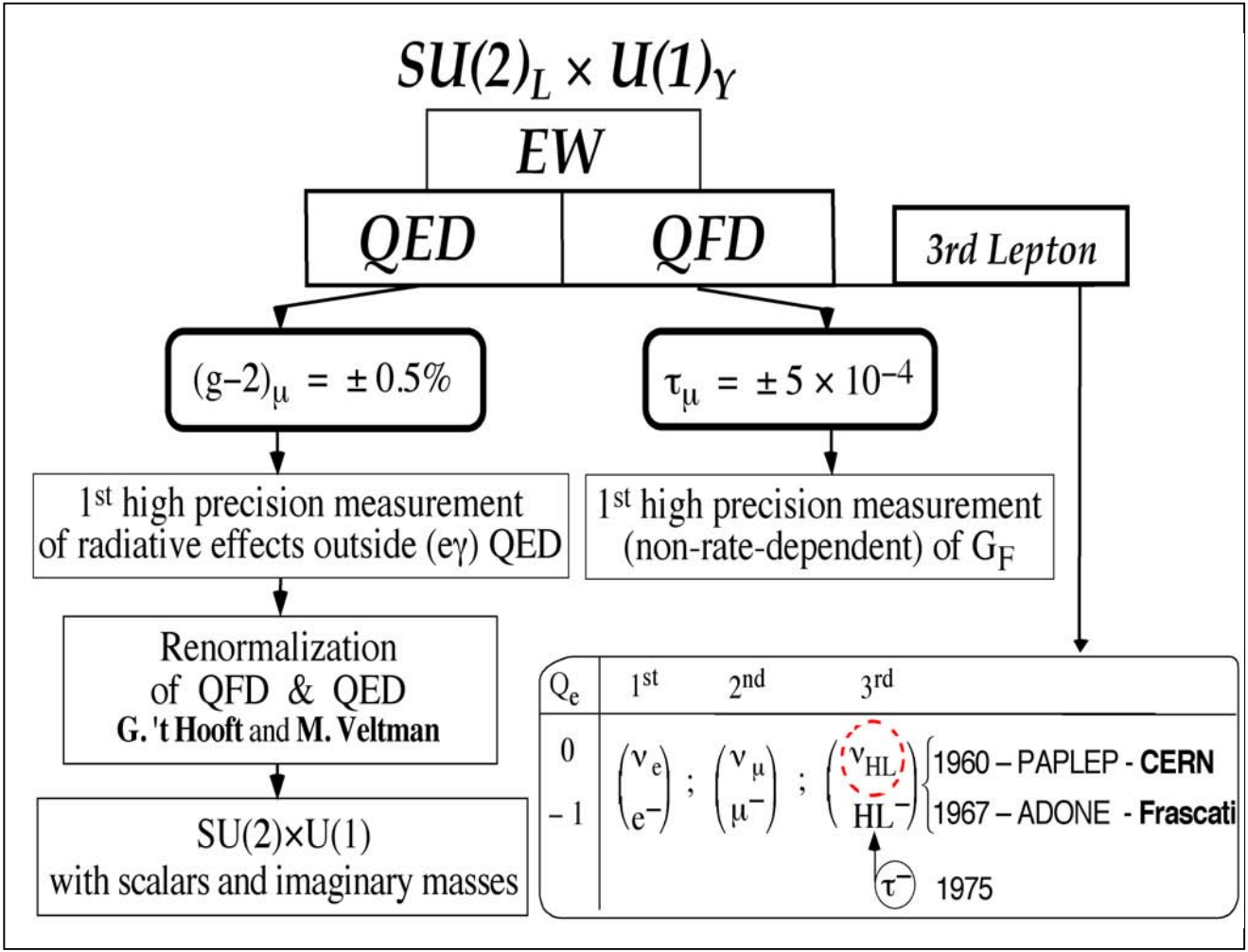

*Fig. 6.*

If the third lepton would have been discovered at CERN we would not have used the symbol τ for the very simple reason that the τ and θ were two mesons decaying into 3 and 2 pions respectively and you cannot use the same symbol to indicate another particle. Furthermore these two mesons, τ and θ, gave rise to the famous **τ – θ puzzle** discovered by Richard Dalitz [5].

All this was neglected by those who, instead of using for the 3$^{rd}$ lepton the (HL$^\pm$) and for the neutrinos ($\nu_{HL}, \bar{\nu}_{HL}$) introduced the notation τ, and $\nu_\tau$ : i.e.

$$\mathbf{HL} \Longrightarrow \boldsymbol{\tau}$$
$$\boldsymbol{\nu}_{\mathbf{HL}} \Longrightarrow \boldsymbol{\nu}_{\boldsymbol{\tau}}\,.$$

The (τ – θ) puzzle was not a trivial detail; it produced the breaking of Parity and of Charge Conjugation Invariance in 1954 thanks to T.D. Lee and C.N. Yang [6]. Dick Dalitz, the author of the (τ – θ) puzzle, regretted very much that the existence of the large time-like (EMFF) of the nucleon did not allow the discovery of the 3$^{rd}$ lepton at CERN in the early sixties. For the very simple reason that, in this case as often pointed out by Dalitz, the notation would not be $\tau$ and $\nu_\tau$ but HL and $\nu_{HL}$. This is why the 44$^{th}$ course of the Subnuclear Physics School in Erice was dedicated to Richard Dalitz, as shown in Figure 7.

*Fig. 7.*

Talking about new ideas, since there are many young physicists in the audience, it is appropriate to recall what Isidor I. Rabi, the founder of the formidable School of Physics in USA, the Physics Department of Columbia University (NY) said: *«Physics needs new ideas. But to have a new idea is a very difficult task: it does not mean to write a few lines in a paper. If you want to be the father of a new idea, you should fully devote your intellectual energy to understand all details and to work out the best way in order to put the new idea under experimental test. This can take years of work. You should not give up. If you believe that your new idea is a good one, you should work hard and never be afraid to reach the point where a new-comer can, with little effort, find the result you have been working, for so many years, to get. The new-comer can never take away from you the privilege of having been the first to* *open a new field with your intelligence, imagination and hard work. Do not be afraid to encourage others to pursue your dream. If it becomes real, the community will never forget that you have been the first to open the field.»* (1972).

This is reproduced in a bronze sculpture placed in the I.I. Rabi Institute of the EMFCSC in Erice (Figure 8). The Ceremony was chaired by Tsung Dao Lee (Figure 9).

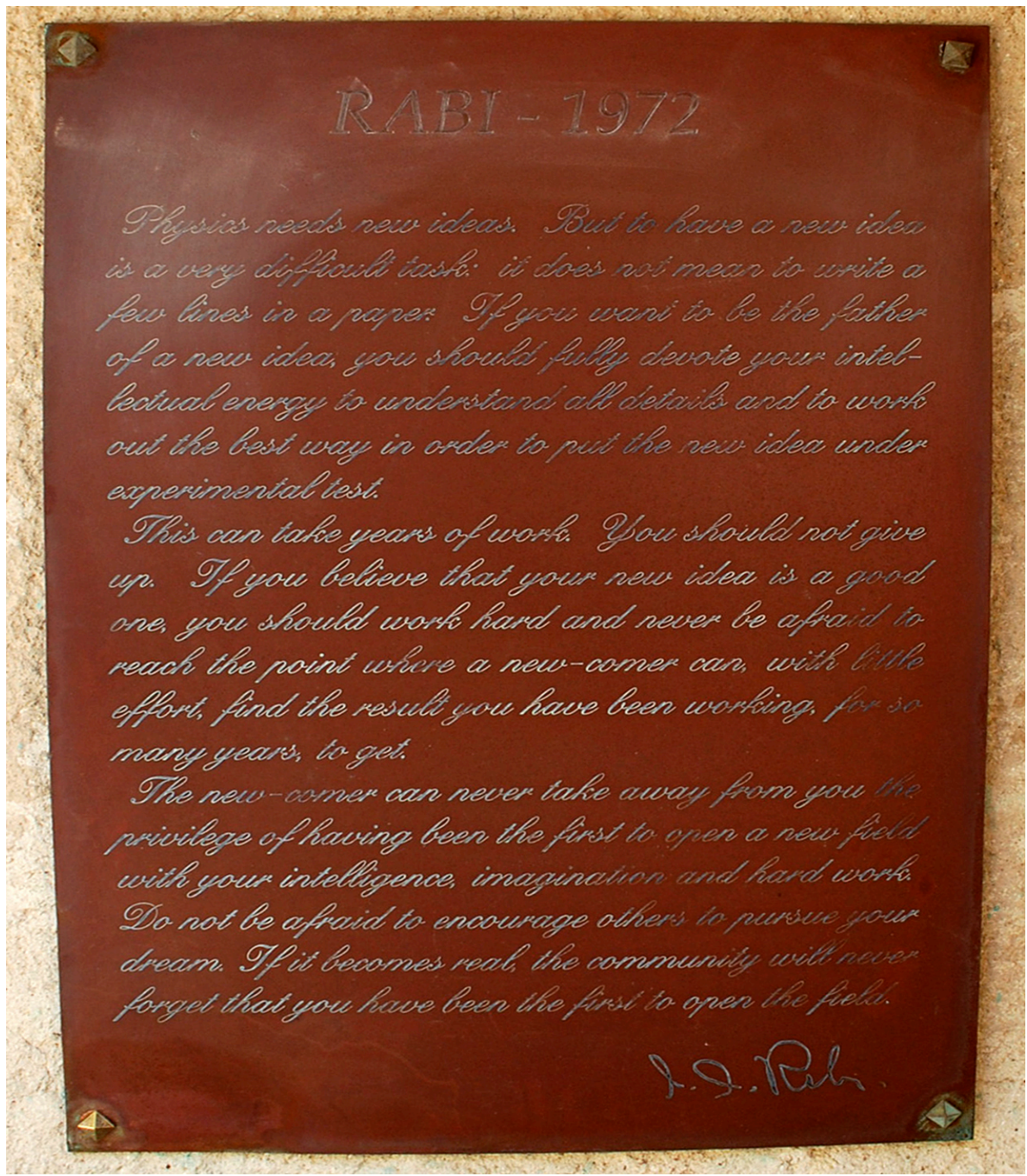

*Fig. 8.*

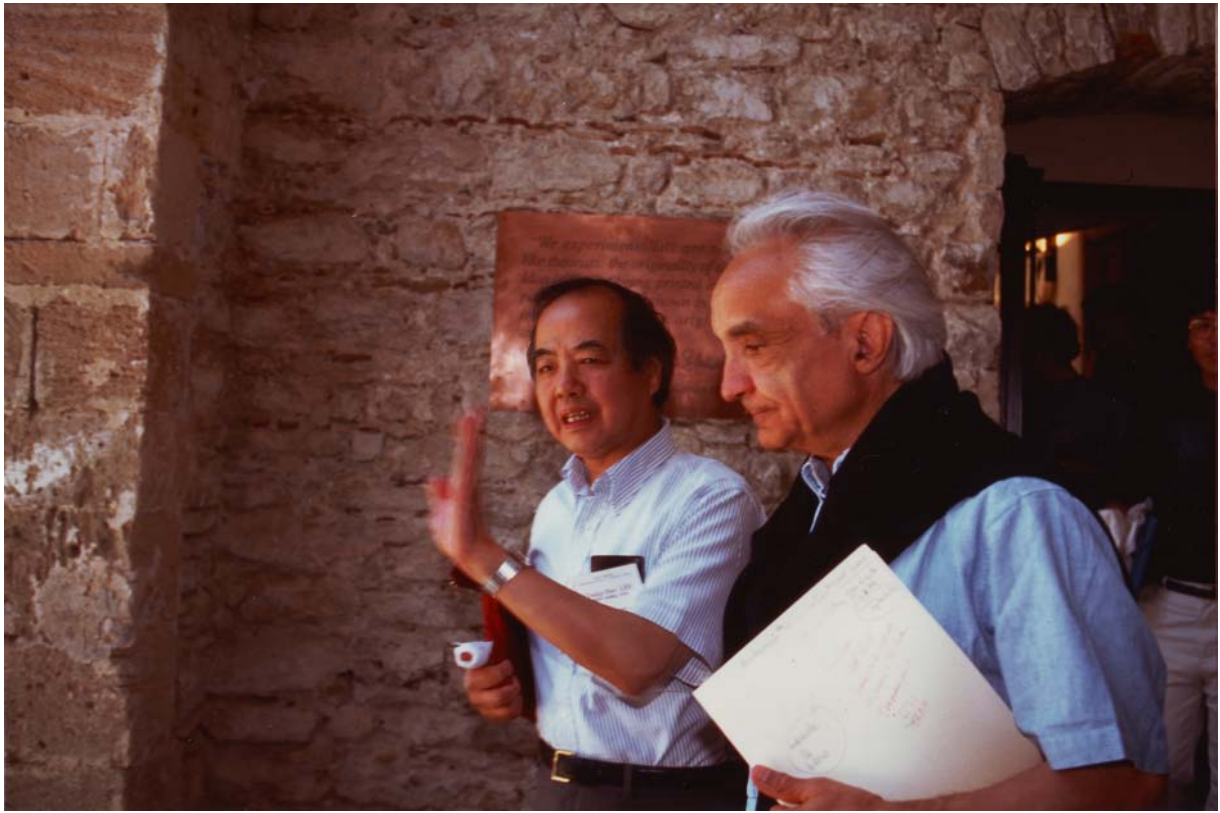

*Fig. 9.*

As you all know the physics of OPERA is focused to the study of neutrino oscillations which – as mentioned before – was indeed in the scientific aims of the Gran Sasso Project (Figure 10) presented by me at the Commission on Public Works of the Italian Senate.

To summarize, the scientific aims of the "Gran Sasso" laboratory are the study of:

1) nuclear stability;
2) neutrino astrophysics;
3) new cosmic phenomenology;
4) **neutrino oscillations;**
5) biologically active matter;
6) ground stability.

*Not only $\tau_p \neq \infty$*

*Fig. 10: Reproduction of page 13 of the original project [7].*

There are two Figures (11 and 12) which I presented at the Commission in 1979, after having discussed with the President of the Italian Republic, Sandro Pertini, the new projects for the future activities of the INFN.

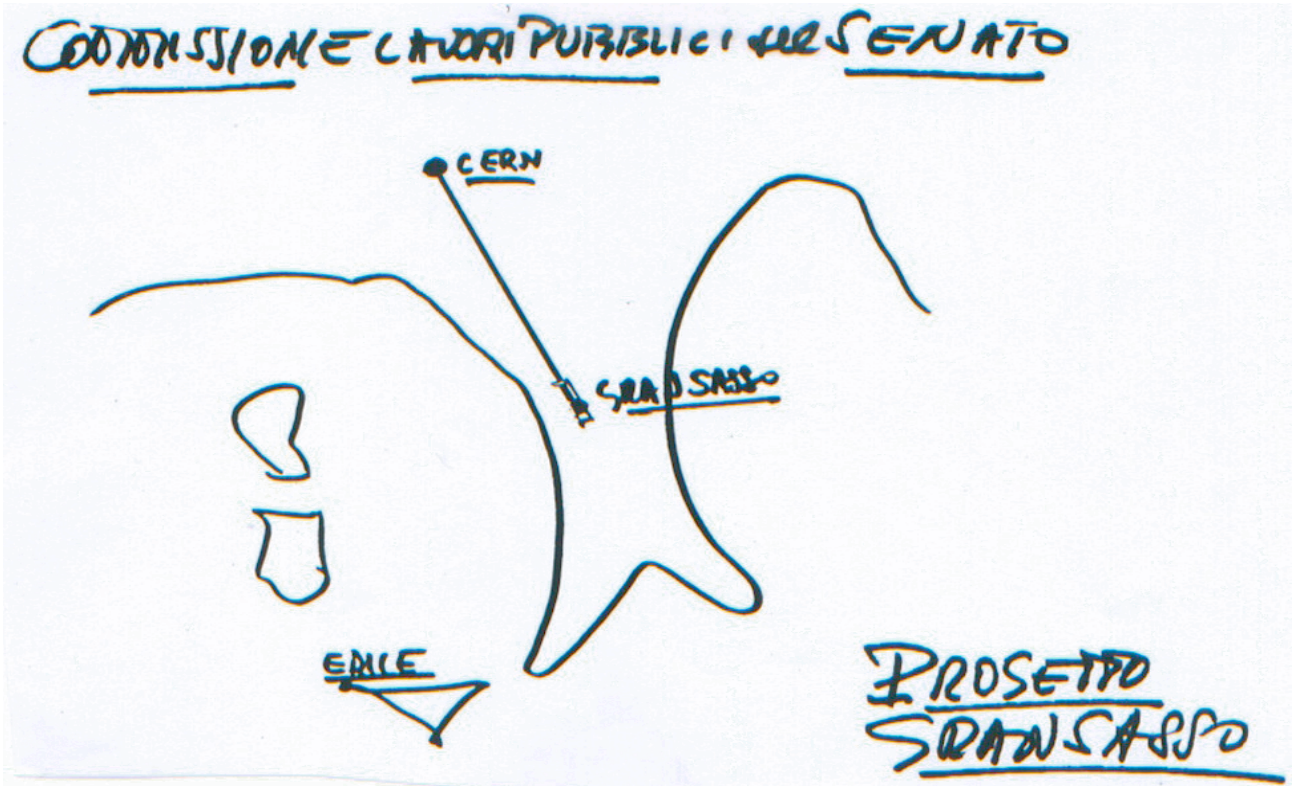

*Fig. 11: Handwritten notes from the presentation by A. Zichichi to the Commission on Public Works of the Senate in a Session convened on short notice by the President of the Senate to discuss the proposal for Gran Sasso Project (1979).*

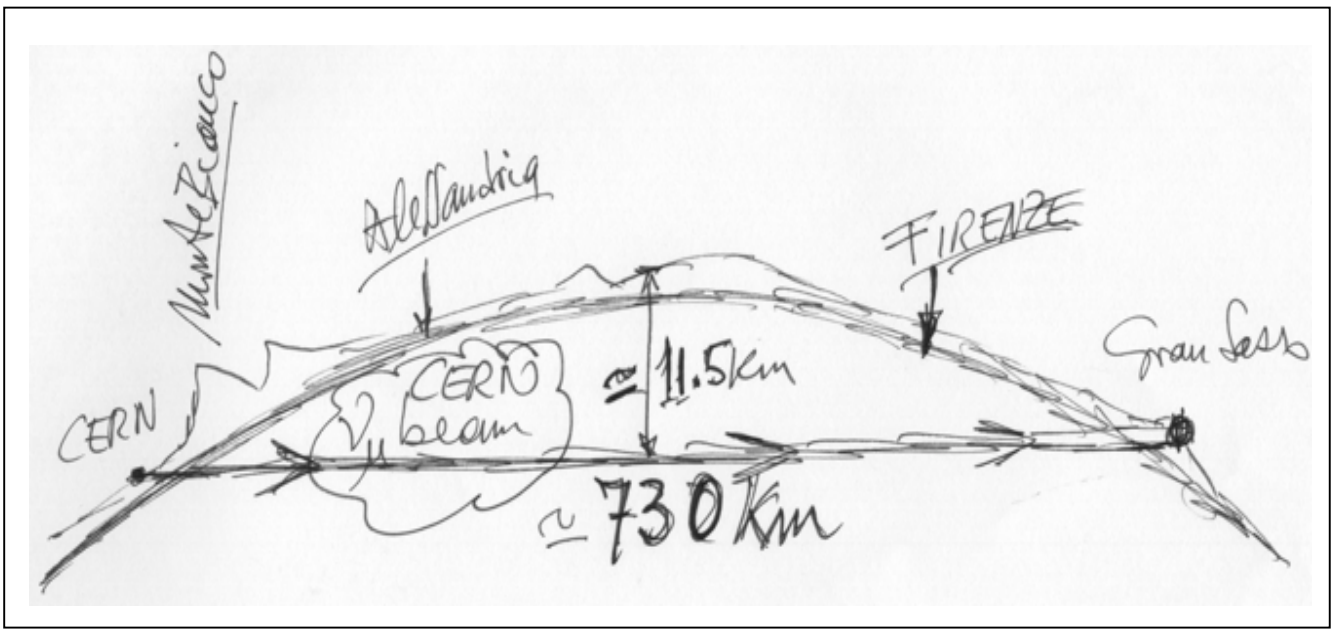

*Fig. 12.*

The two Figures (11 and 12) produced a great interest in our President, who became a strong supporter of the INFN five-years plan where the proposal was to have an increase of the budget by an order of magnitude: from 20 billion lira to 200 billion. My friends were saying Nino is dreaming; the non-friends were using less gentle expressions.

I want to recall that, for the first time in the history of INFN, the presentation was at the special Hall of the Italian Parliament and in the first rank there was the President of the Italian Republic, Sandro Pertini with many members of the Government.

Nevertheless many physicists, who Fermi would have classified of rank 2 and 3, were continuing their campaign against the Gran Sasso Project.

Fortunately Bruno Pontecorvo (Figure 13) was suddenly allowed to visit Italy.

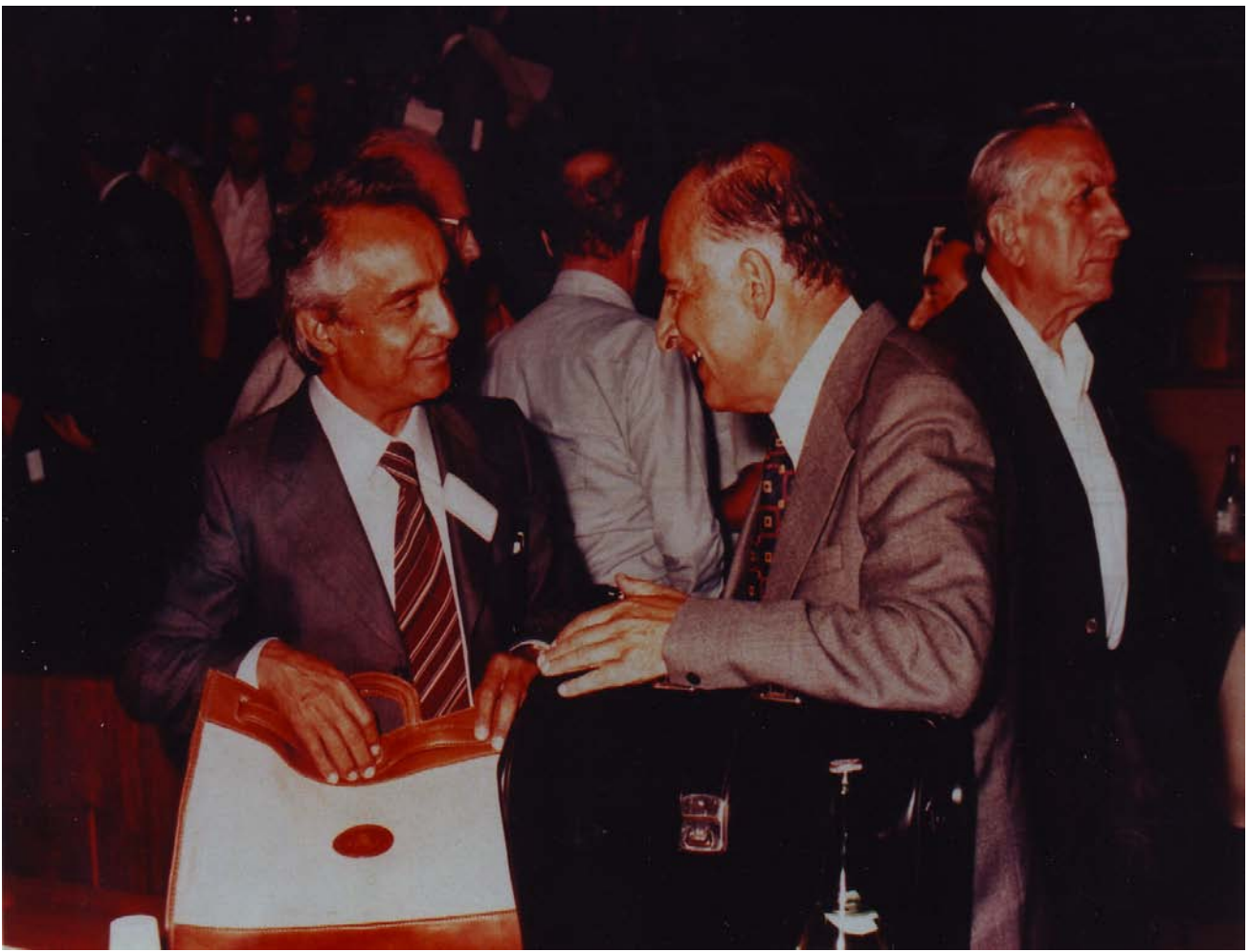

*Fig. 13: The author and Bruno Pontecorvo, Rome, September 1978.*

The Berlin Wall was up and this became a great occasion for the media. It was the time when the Gran Sasso Project was being discussed in the scientific environment.

The so-called "Rome School" tried to bring to a halt the "Gran Sasso Project" with arguments similar to those used at Frascati with ADONE to boycott the energy increase of the ($e^+e^-$) collider («Zichichi is searching for butterflies»), when I was searching for the 3rd lepton after PAPLEP using the ($e^+e^-$) annihilation many years before Perl.

Underground lobbying by the "Rome-School" had produced effects. For instance the CERN-Director General (DG) Leon Van Hove, during a CERN Council meeting declared that Zichichi's Gran Sasso Project was invented to stop the joint venture between Italy and France to realize an underground Laboratory in the Fréjus tunnel. After this unprecedented attack against a very important initiative in Italy, the other CERN-DG, John Adams, called me into his office, to tell him «not to worry». This ended all attacks from CERN against the project.

The arrival of Pontecorvo in Italy gave rise to a new set of actions. During the visit and the various lectures by Pontecorvo, a journalist asked: «Professor Pontecorvo, what do you think of the Gran Sasso Project proposed by Professor Zichichi? Many physicists consider it a useless Napoleonic venture with weak scientific content». After a few seconds of thinking, in the usual Pontecorvo style of soft and slow answering, he said: «I regret not to be young enough to participate in this formidable project. The scientific content of the project appears to me extremely interesting».

This declaration by Pontecorvo came as a surprise, since we were on opposite political sides and every journalist was expecting a strong negative statement from Pontecorvo on the Gran Sasso Project. But physics prevailed. And this put an end to all — open as well as underground — lobbying against the Gran Sasso Project.

Bruno Pontecorvo never lost any occasion to emphasize the value of studying the oscillations between neutrino flavours. In fact in his original paper he was suggesting the study of ($\nu\bar{\nu}$) oscillations. The flavours oscillations came later, as reported in a note by Gillo Pontecorvo (Figure 14).

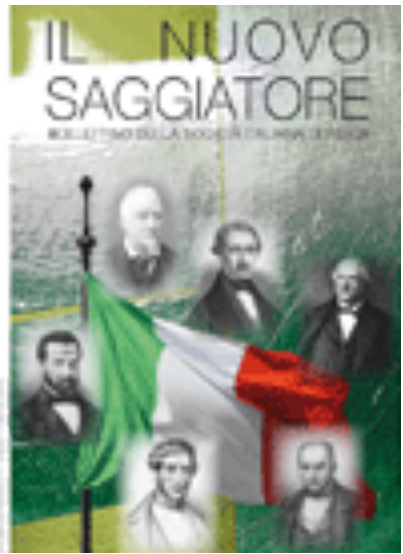

**NEUTRINO MIXING AND BRUNO PONTECORVO**

I have been reading with great pleasure the beautiful volume edited by SIF: Thirty Years of Gran Sasso, in honour of Antonino Zichichi. This volume contains an impressive series of articles and recollections about the Gran Sasso underground laboratory of INFN, where neutrino physics has always been and still is the main line of research. The results from the BOREXINO experiment and those from the CNGS (CERN to Gran Sasso) neutrino-beam OPERA experiment, together with the start of the ICARUS experiment, have newly captured the attention of the physics community the world over and the problem of neutrino mixing has again come to the limelight.

Recently, on the major Italian daily press, the credit for the mixing of the 3rd-family neutrino with neutrinos from the other two families has been attributed to my father Bruno Pontecorvo. In fact his original idea, back to the late Fifties, when only one neutrino (ne) was known, concerned neutrino-antineutrino oscillation and became neutrino-flavour mixing a few years later when the second neutrino (nμ) was discovered.

It was in 1964, at the Dubna Conference, that Zichichi presented the first results of his PAPLEP experiment at CERN to search for a 3rd heavy lepton with its own neutrino by detecting acoplanar electron-muon pairs. He discussed his experiment with Bruno who, I remember, already at that time was impressed by Zichichi's idea. This idea is at the origin of the CERN to Gran Sasso neutrino beam, included in the original Gran Sasso Project that Zichichi presented to the Italian Government in 1979, as duly reported in Thirty Years of Gran Sasso. To study the oscillation of the 3rd-family neutrino into neutrinos from the other two families was first proposed by Nino Zichichi, not by Bruno who, in many occasions along the years, expressed his strong support to the Gran Sasso Project and its author.

Gil Pontecorvo
JINR, Dubna, Russia

OPINIONI Online First
4 agosto 2011

*Fig. 14.*

Let me close this brief recollection of the origin of the LNGS showing the basic characteristics of the Lab (Figure 15), the cover page (Figure 16) of a volume edited by Sandro Bettini, on the occasion of the XX Anniversary of the Gran Sasso Lab and a synthesis of the evolution of the three gauge couplings (Figure 17). This evolution was studied in order to illustrate the complementarity of the Gran Sasso Project and the other projects (LEP, LHC and ELN) where high energy colliders were involved. The complementarity of all projects proposed in the five-years INFN plan (where the increase by an order of magnitude for the financial support was requested) had to be proved with strong physics arguments in order to convince 1$^{st}$ rank fellows.

## 2. THE BASIC CHARACTERISTICS OF THE LAB.

The range of scientific perspectives opened up by the Gran Sasso Laboratory goes far beyond the measurement of the proton lifetime, as shown in Fig. 1.1.

These *scientific perspectives* depend on the *basic features* of the Gran Sasso Laboratory, which are:

1) very low noise due to local radioactivity;
2) neither too deep, nor too shallow underground;
3) orientation towards the most powerful (artificial) source of neutrinos and other unknowns (Fig. 2.1);
4) link with a laboratory at the top of the Gran Sasso, which allows time coincidences to be made (Fig. 2.2);
5) instrumentation which uses the most advanced technologies.

The *low noise* level in terms of *natural radioactivity*, was proved before the excavation work started. The measurements of the cosmic ray flux and of the local rock radioactivity were first performed by one of my collaborators, – L. Federici [3] – whom I want to pay tribute to, in this solemn occasion. These measurements demonstrated that over the length of one Km the cosmic ray flux was constant. This nice feature is due to the shape and structure of the mountain. The Gran Sasso rock radioactivity was so low that the term «laboratory of cosmic silence» could be coined.

### *REFERENCES*

[1a] *THE GRAN SASSO PROJECT.* A. Zichichi, Proceedings of the GUD-Workshop on "Physics and Astrophysics with a Multikiloton Modular Underground Track Detector", Rome, Italy, 29-31 October 1981 (INFN, Frascati, 1982), 141.

[1b] *THE GRAN SASSO PROJECT.* A. Zichichi, NFN/AE-82/l, 28 February 1982.

[1c] *THE GRAN SASSO PROJECT.* A. Zichichi, Proceedings of the Workshop on "Science Underground", Los Alamos, NM, USA, 27 September-1 October 1982 (AIP, New York, 1983), 52.

[2] *THE GRAN SASSO LABORATORY.* A. Zichichi, Proceedings of the International Colloquium on "Matter Non-Conservation" - ICOMAN '83, Frascati, Italy, 17-21 January 1983 (INFN, Frascati, 1983), 3.

[3] *THE GRAN SASSO LABORATORY AND THE ELOISATRON PROJECT.* A. Zichichi, "Old and New Forces of Nature" (Plenum Press, New York-London, 1988), 335.

[4] *PERSPECTIVES OF UNDERGROUND PHYSICS: THE GRAN SASSO PROJECT.* A. Zichichi, Invited Plenary Lecture at the Symposium on "Present Trends, Concepts and Instruments of Particle Physics", in honour of Marcello Conversi's 70th birthday, Rome, Italy, 3-4 November 1987, *Conference Proceeding*s Eds. G. Baroni, L. Maiani and G. Salvini (SIF, Bologna, 1988), Vol. 15, 107.

*Fig. 15: Reproduction of page 111 of Ref. 8; Proceedings in honour of M. Conversi, a strong supporter of the Gran Sasso Project.*

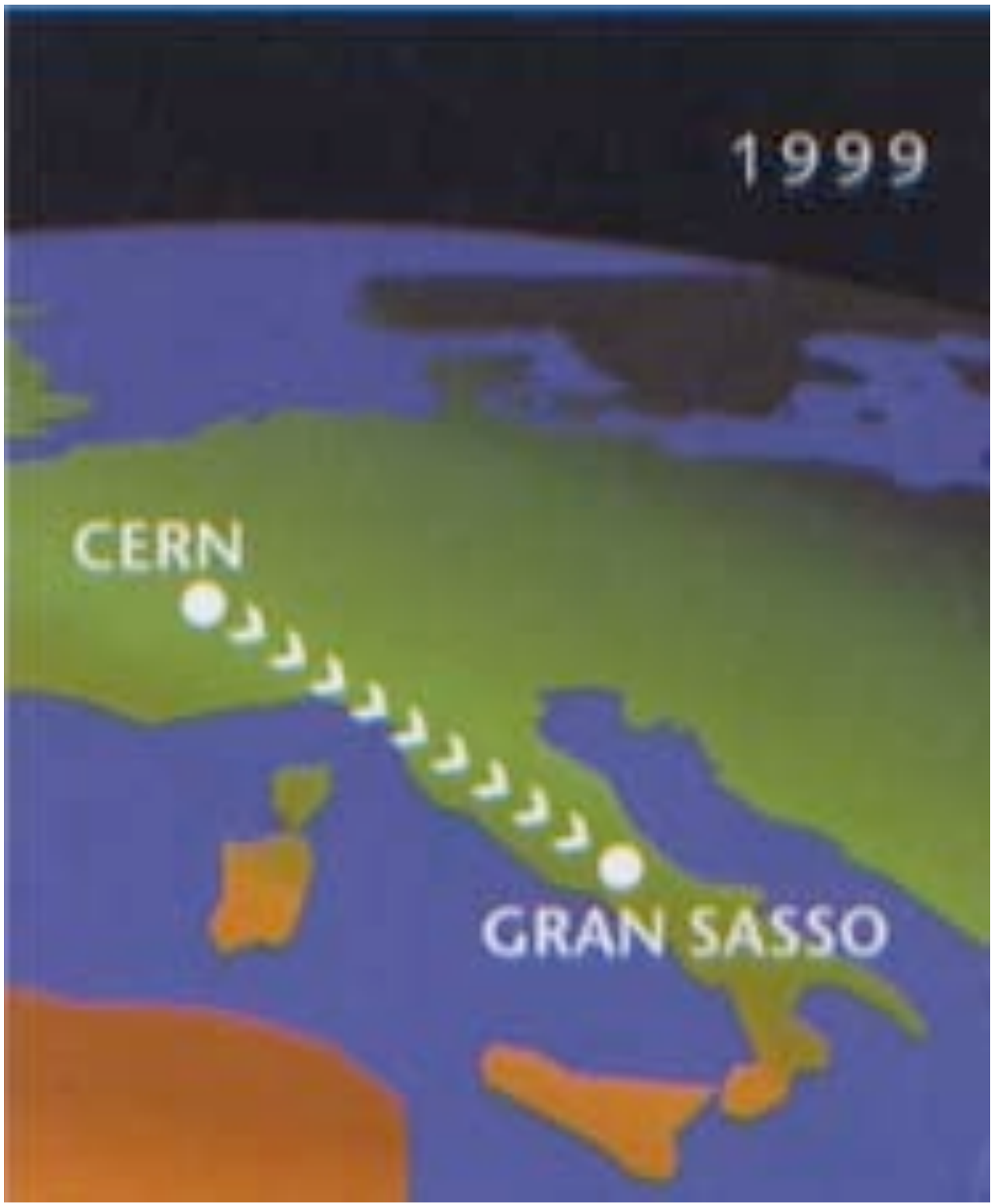

*Fig. 16.*

*Fig. 17: This picture shows the pre-Big Bang zone, which is at the centre of theoretical attention today.*

This is because, during the difficult times of the Gran Sasso Projects there were 2[nd] and 3[rd] rank people called physicist doing underground lobbying and stating that this project was against the other Italian engagements in Europe which I had included in the five-years INFN plan.

In fact in addition to the Gran Sasso Project, I presented other two important engagements for our Country, LEP in Geneva and HERA in Hamburg.

Let me recall Enrico Fermi Statement where the meaning of the 2[nd] and 3[rd] rank fellows is given: «*There are several categories of scientists in the world; those of second or third rank do their best but never get very far. Then there is the first rank, those who make important discoveries, fundamental to scientific progress. But then there are the geniuses, like Galilei and Newton. Majorana was one of these*» (Rome 1938).

In order to convince 1[st] rank fellows you need strong arguments [9], such as the Renormalization Group Equations, originated in 1951 by my friend André Petermann who was the pupil of Stueckelberg [10].

The mathematical formalism which has been used to obtain the results shown in Figure 17 is a system of three differential non-linear equations coupled via the gauge couplings

$$\alpha_i, \; \alpha_j \; (\text{with } i = 1, 2, 3; \text{ and } J = 1, 2, 3 \text{ but } i \neq j),$$

as shown in Figure 18.

---

**THE UNIFICATION OF ALL FUNDAMENTAL FORCES**

$$\mu \frac{d\alpha_i}{d\mu} = \frac{b_i}{2\pi} \alpha_i^2 + \sum_j \frac{b_{ij}}{8\pi^2} \alpha_i \alpha_j$$

The three lines in Figure 17 result from calculations executed with a supercomputer using the system shown above.

---

*Fig. 18.*

During more than ten years (from 1979 to 1991), no one had realized that the energy threshold for the existence of the Superworld was strongly dependent on the 'running' of the masses [11].

This is now called: the EGM effect (from the initials of Evolution of Gaugino Masses). To compute the energy threshold using only the 'running' of the gauge couplings ($\alpha_1$, $\alpha_2$, $\alpha_3$) corresponds to neglecting nearly three orders of magnitude in the energy threshold for the discovery of the first particle (the lightest) of the Superworld [12-28].

The reason why I have devoted some attention to all these past details is in another Fermi Statements: *«Neither Science nor civilization could exist without Memory»*.

If we are here today, if the Gran Sasso Lab exist, if LVD and OPERA collaborate, if all this can happen, this is because all the difficulties mentioned before have been passed. The difficulty we are facing today is just an example of the problems which need to be overcome having as primary purpose to do good physics.

Good Physics was (and is) the reason why OPERA has been built: i.e. to study the oscillation between the second and the third neutrinos.

Since these are times when the physics community is interested in good and bad behaviour in physics research I thought it was appropriate to recall that, when the Gran Sasso Project was elaborated, the CERN-Gran Sasso neutrino beam was proposed as a good experiment to be done in order to study the oscillations between the muon neutrinos $\nu_\mu$ and the $\nu_{HL}$. Let us not loose the memory, as Enrico Fermi wanted.

I will now say few words on the LVD and OPERA detectors and then discuss the results obtained by the collaboration between the two groups, LVD and OPERA, in order to establish if a time-shift occurred during the calendar years (1997–2012).

## 4  THE LVD DETECTOR

The Large Volume Detector (LVD) [29] (Figure 19) is located in Hall A of the INFN underground Gran Sasso National Laboratory at an average depth of 3600 *m. w. e.* The LVD main purpose is to detect and study neutrino bursts from galactic gravitational stellar collapses.

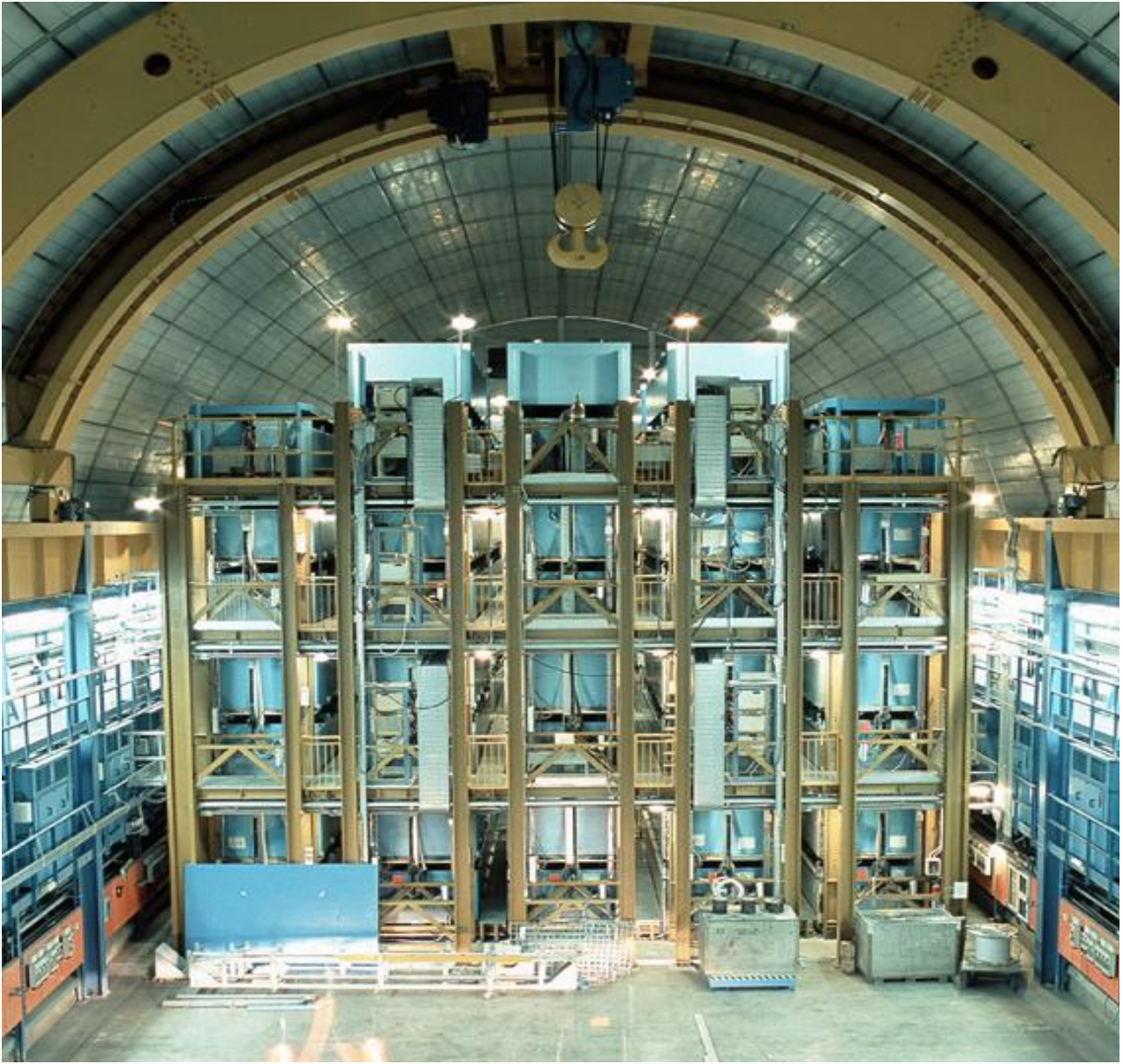

*Fig. 19: Photograph of the LVD detector in the Hall A of the underground Gran Sasso Labs.*

The experiment started taking data in June 1992 and has continued without interruptions until now. The detector, schematically shown in Figure 20, consists of an array of 840 liquid scintillator counters, 1.5 m$^3$ each, arranged in a compact and modular geometry: 8 counters are assembled in a module called "portatank"; 35 portatanks (5 columns × 7 levels) form a "tower"; the whole detector consists of three identical towers that have independent power supply, trigger and data acquisition systems.

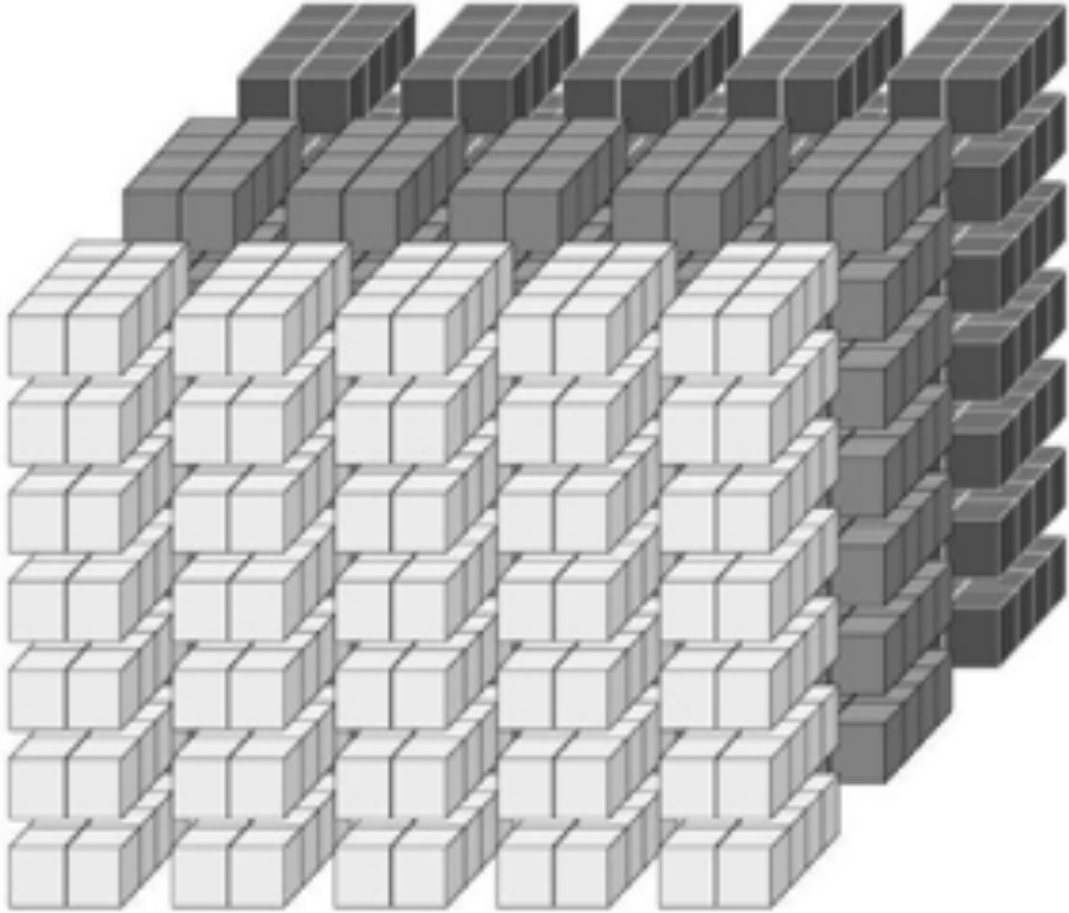

*Fig. 20: Schematic view of the LVD apparatus.*

The external dimensions of the active volume are $13 \times 23 \times 10$ m$^3$. The liquid scintillator (density $\rho = 0.8$ g/cm$^3$) is $C_nH_{2n}$ with $<n> = 9.6$ doped with 1 g/l of PPO (scintillation activator) and 0.03 g/l of POPOP (wavelength shifter). The total active scintillator mass is M = 1000 t. Each LVD counter is viewed from the top by three 15 cm diameter photomultipliers (FEU49 or FEU125). The main reaction that is detected by LVD is the inverse beta decay

$$(\bar{\nu}_e \, p, \, ne^+),$$

which gives two signals: a prompt one due to the e$^+$ followed by the signal from the neutron capture reaction (np,d$\gamma$) with mean capture time of about 185 µs and $E_\gamma = 2.2$ MeV.

The modularity of the apparatus allows for calibration, maintenance and repair interventions without major negative interference with data taking and detector sensitivity. Figure 21 shows the duty cycle and the trigger active mass of LVD from June 1992 to March 2011.

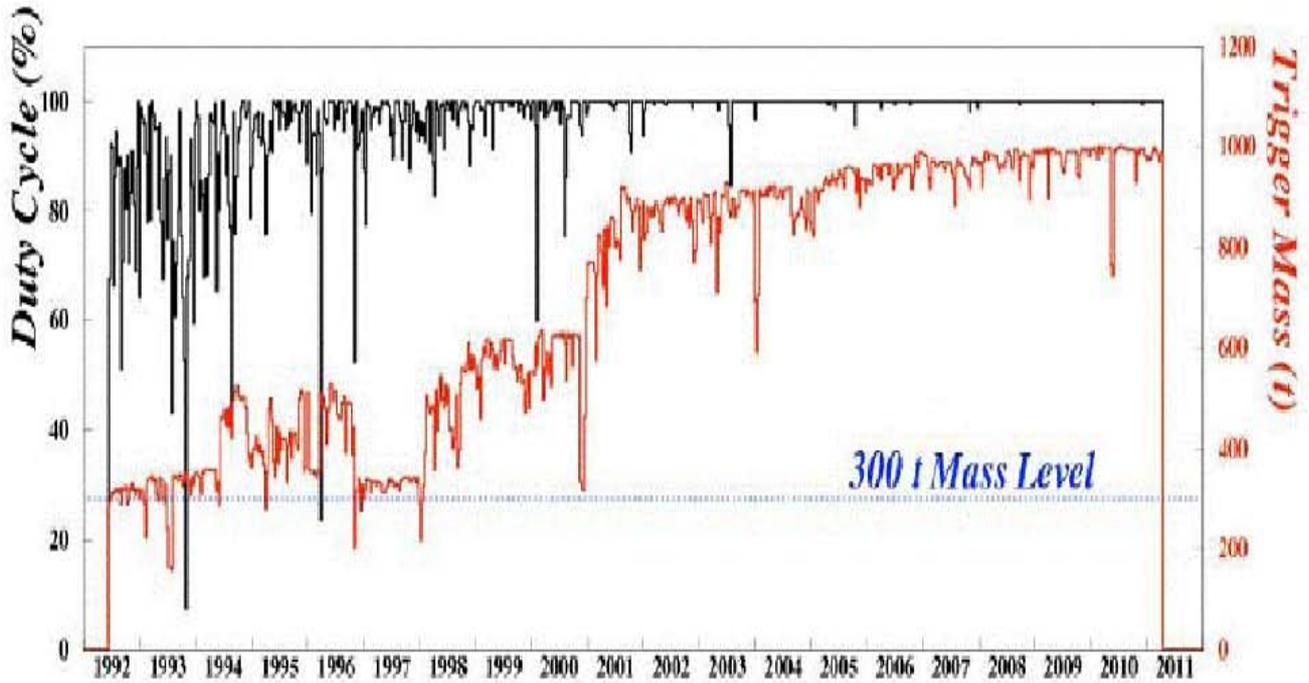

*Fig. 21: LVD duty cycle and active mass in the period June 1992 – March 2011.*

From 2001 the experiment has been in very stable conditions with duty cycle > 99% and slightly increasing active mass.

The minimum trigger mass of 300 t, corresponding to less than one "tower", at which LVD can monitor the whole Galaxy for gravitational core collapses is also shown (blue).

## 5 THE OPERA DETECTOR

OPERA is a hybrid experiment with electronic detectors and nuclear emulsions located in Hall C of the underground Gran Sasso Laboratory [30]. The main physics goal of the experiment is to observe neutrino flavour oscillations through the appearance of $v_\tau$ via the production of a $\tau$ lepton in the $v_\mu$ CNGS beam.

The detector design was optimized to identify the $\tau$ lepton via the topological observation of its decay: this requires a target mass of more than a kton to maximize the neutrino interaction probability and a micrometric resolution to detect the $\tau$ decay.

# 6 THE TIME DIFFERENCE BETWEEN LVD AND OPERA

The time difference between LVD and OPERA is given by

$$\delta t = t_{LVD} - t_{OPERA}$$

In order to study the stability of the time difference δt versus calendar time, the data have been subdivided in different periods of the various solar years. The year 2008 has been divided in three samples: before May, May-August, after August. For each period we look at the δt distribution, compute the mean and the RMS.

The results are shown in Figures 22 and 23 and summarized in Table 1. The total number of events, 306, is distributed into eight samples, each one covering a given calendar time period.

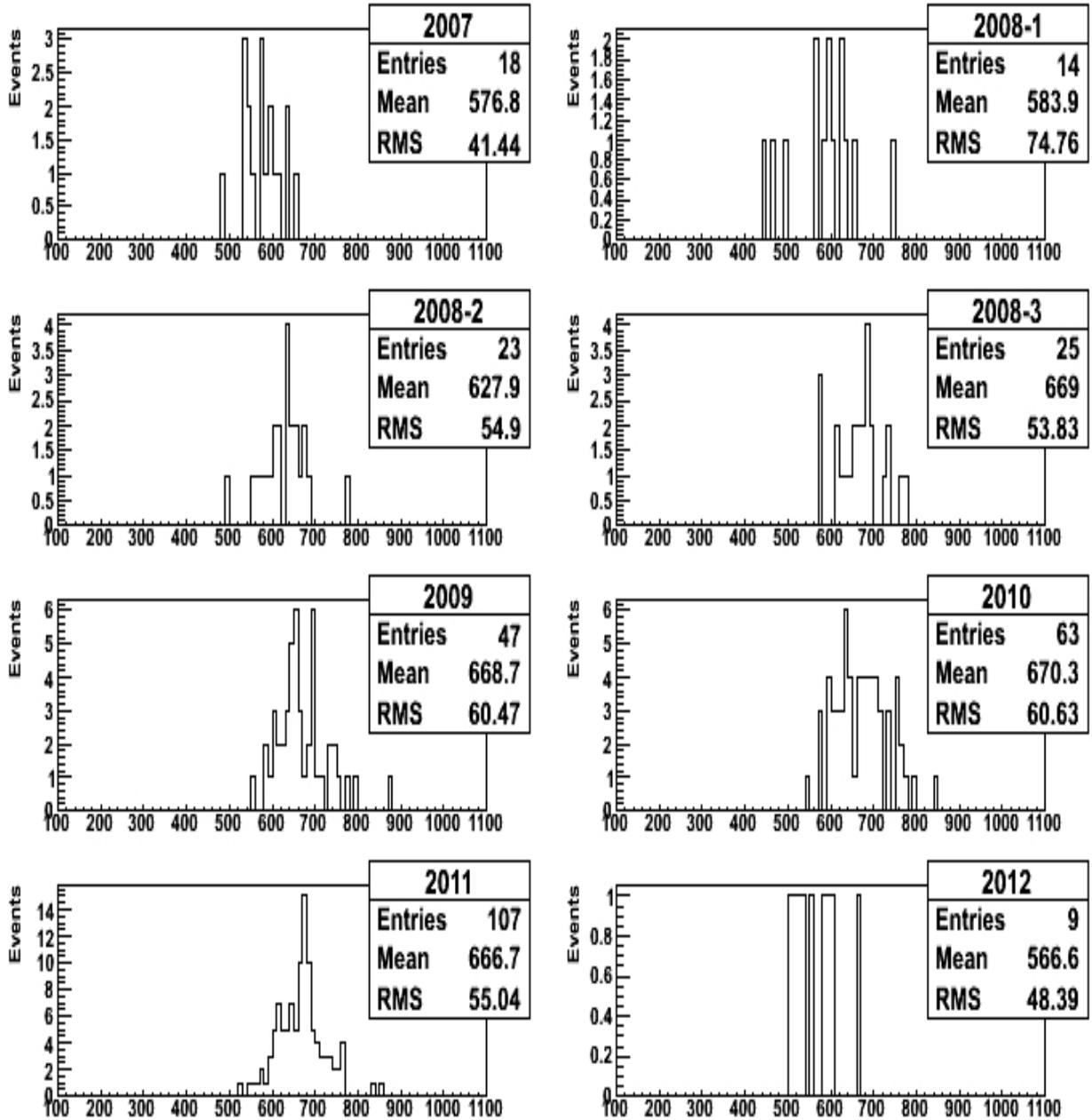

Fig. 22: Distribution of the $\delta t = t_{LVD} - t_{OPERA}$ for each period of time. In the vertical axis the number of events; in the horizontal axis δt in ns.

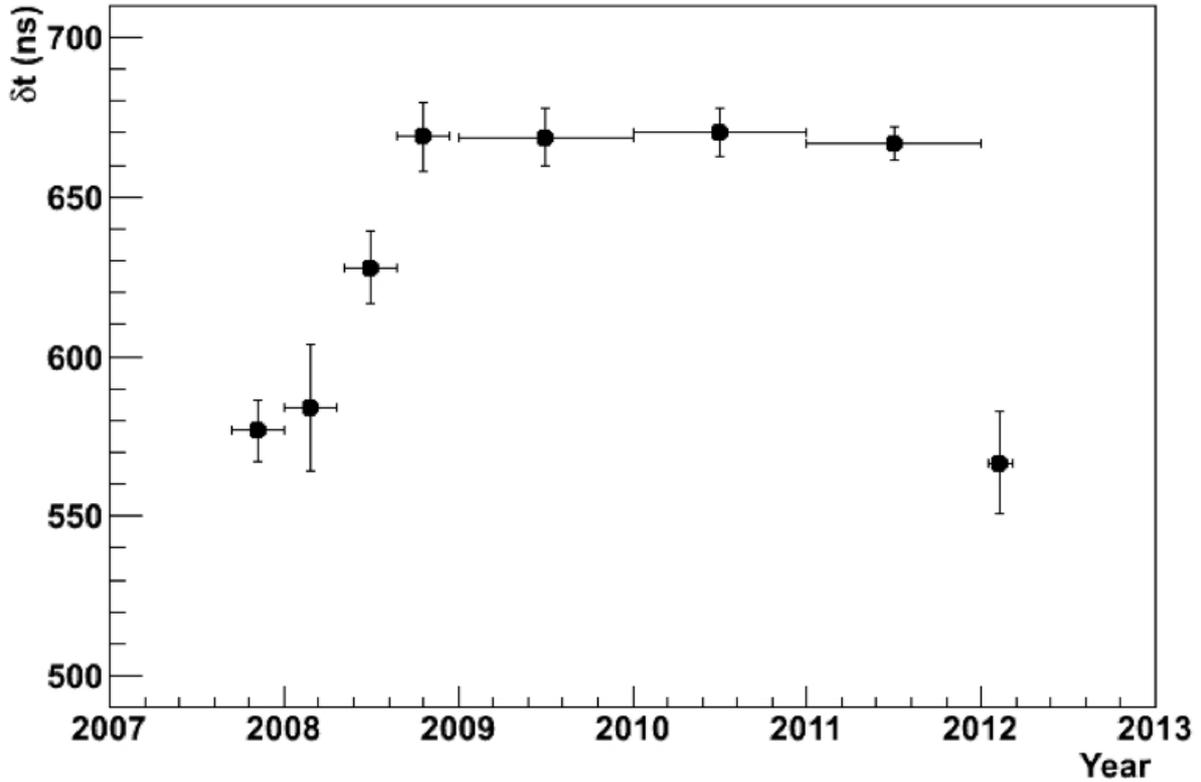

*Fig. 23: Distribution of the δt = $t_{LVD} - t_{OPERA}$ for corrected events. All the events of each year are grouped in one single point with the exception of year **2008** which is subdivided in three periods: before May, May-August, after August.*

| \multicolumn{6}{c}{**TOTAL NUMBER OF EVENTS = 306**} |||||| 
|---|---|---|---|---|---|
| **Class** | **Year** | **Since** | **To** | **Nb. of events** | **<δt> (ns)** |
| A | 2007 | Aug | Dec | 18 | 577± 10 |
| A | 2008-1 | Jan | Apr | 14 | 584 ± 20 |
| A | 2008-2 | May | Aug | 23 | 628 ± 11 |
| A | 2012 | Jan | Mar | 9 | 567 ± 16 |
| B | 2008-3 | Sep | Dec | 25 | 669 ± 11 |
| B | 2009 | Jun | Nov | 47 | 669 ± 9 |
| B | 2010 | Jan | Dec | 63 | 670 ± 8 |
| B | 2011 | Jan | Dec | 107 | 667 ± 5 |

*Table 1: Summary of the δt distribution in the various calendar time periods.*

Let us now group the results in two classes:

- class A: between August 2007 to August 2008 and from January to March 2012;
- class B: from August 2008 to December 2011.

The two distributions for class A and B are reported in Figure 24.

We obtain for class A

$$\delta t = (595 \pm 8) \text{ ns},$$

while for class B

$$\delta t = (668 \pm 4) \text{ ns}.$$

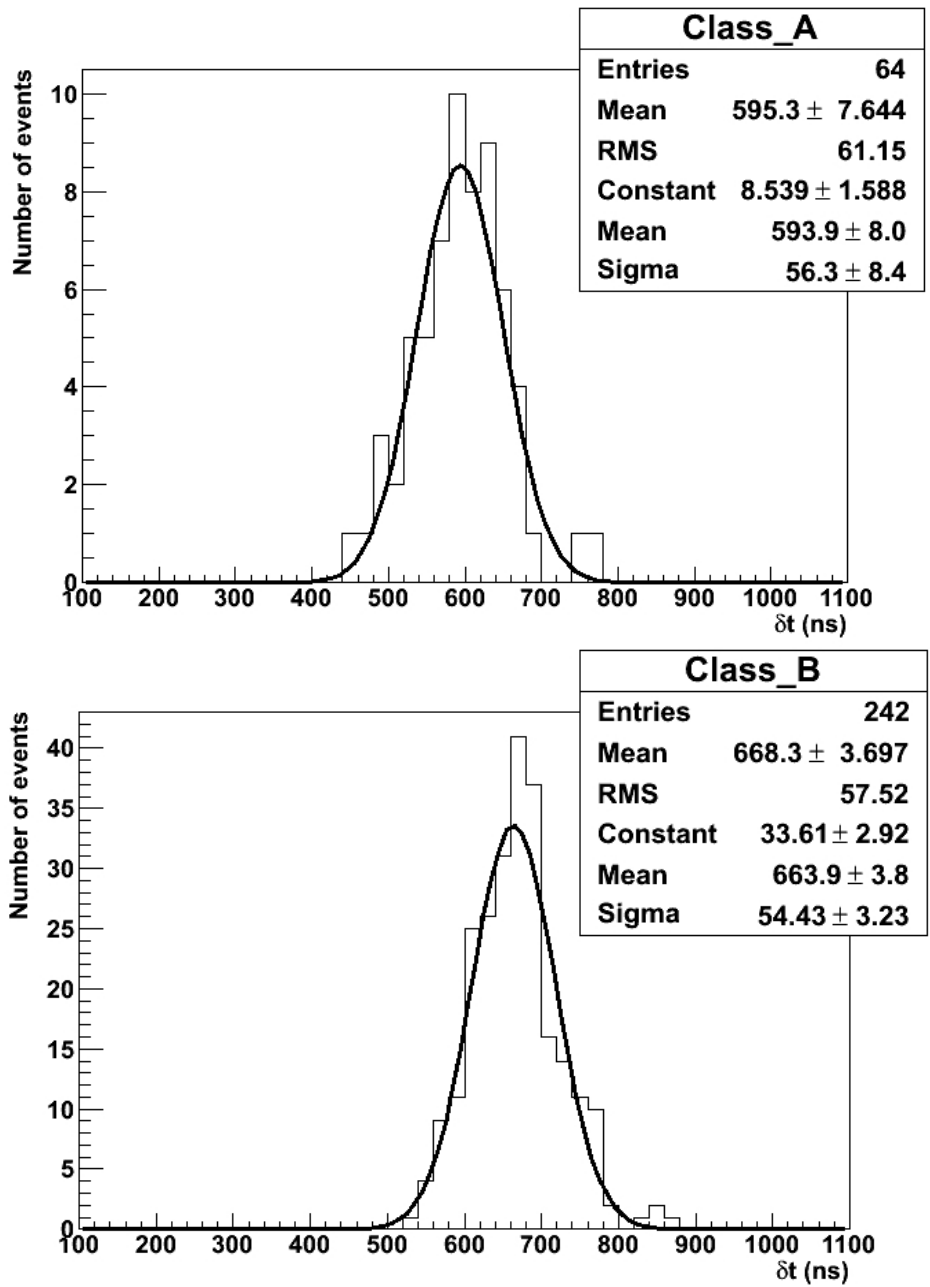

Fig. 24: Distribution of $\delta t = t_{LVD} - t_{OPERA}$ for events of class A and B.

In Figure 25 we report the average value

$$\langle \delta t \rangle$$

for the two classes.

The resulting time difference between the average values in the two classes is

$$\Delta_{AB} = \langle \delta t_A \rangle - \langle \delta t_B \rangle = (-73 \pm 9) \text{ ns},$$

far from zero at 8-sigma level.

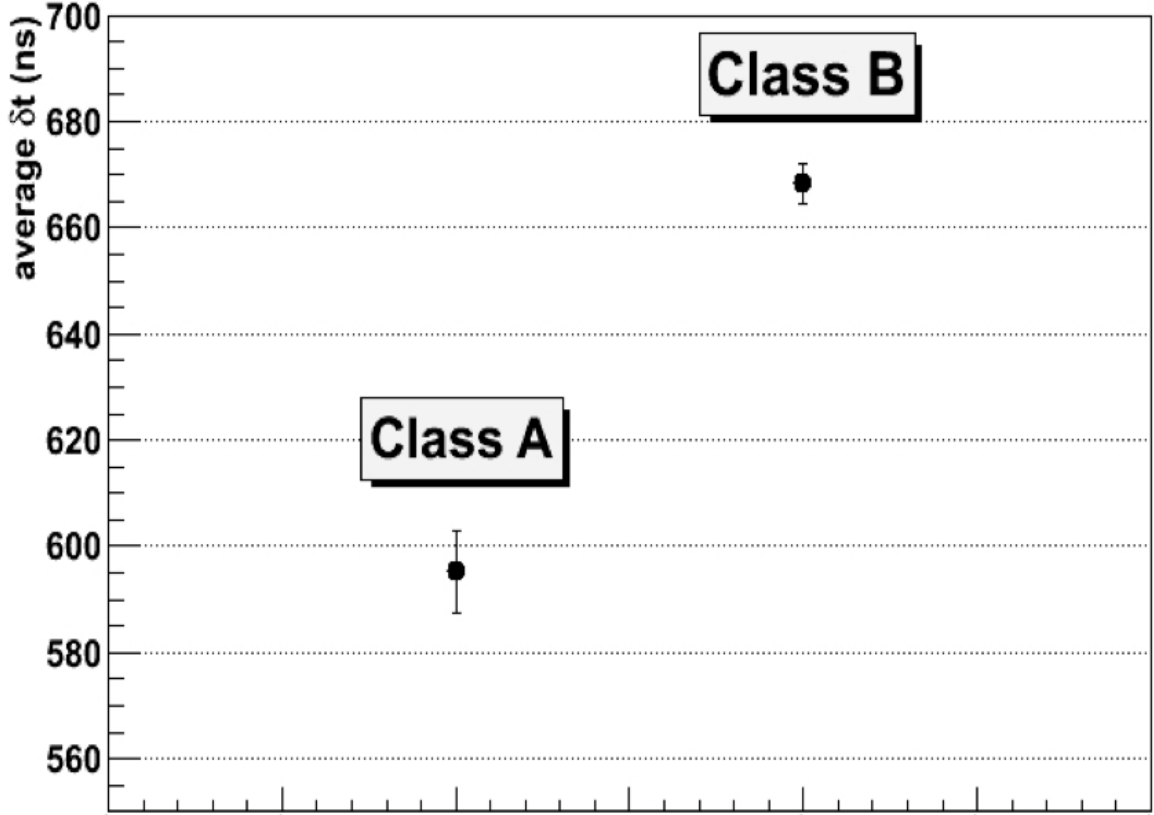

*Fig. 25: Average value of δt computed in each class of events. Class A are events in calendar time from August 2007 to August 2008 and from January 2012 to March 2012; Class B are from August 2008 to December 2011.*

We also note that now, after doing all the needed corrections, the two Gaussian distributions have a width compatible with the ~50 ns time accuracy claimed by the experiments.

The stability in time of LVD shows that the OPERA detector has a negative time shift in the calendar period from August 2008 to December 2011 of the order of

$$\Delta_{AB} = (-73 \pm 9) \text{ ns}$$

compared with the calendar time from August 2007 to August 2008 and from January to March 2012 taken together.

# 7 SUMMARY AND CONCLUSIONS

Data from horizontal muons traversing the LVD and OPERA detectors cover a calendar time period from mid 2007 until 2012, for a total live time of about 1200 days.

In a time-window of 1 μs, and excluding events in time with the CNGS beam spill, we found 306 events due to horizontal muons from the "Teramo anomaly".

This sample has a time-difference ($t_{LVD} - t_{OPERA}$) distribution peaked at 616 ns with an RMS of 74 ns.

The central value of the distribution has the following interpretation: the coincident events detected up to now are not multiple muons (one per each detector), but single muon events entering horizontally from the OPERA side and going through the LVD detector after 616 ns of flight.

Indeed, the OPERA-LVD direction lies along the so-called "Teramo anomaly", where the mountain profile exhibits an anomaly in the *m. w. e.* depth in the horizontal direction. Visual inspection using the event displays of LVD and OPERA detectors confirms this anomaly discovered by LVD in 1997 [1].

The calendar time evolution of the time difference δt for various periods of data-taking is shown in Figure 23.

We see an evolution of the average value in each period, ranging

$$\text{from} \sim 580 \text{ ns in 2007 up to} \sim 670 \text{ ns}$$
$$\text{from May 2008 to the end of 2011;}$$

then for the 9 events collected so far in 2012 it decreases again

$$\text{to} \sim 570 \text{ ns.}$$

The observed variations are larger than the statistical uncertainty estimated for each period.

Grouping the time periods in two classes, as labelled in Table 1, we obtain for class A an average value of

$$\Delta t \text{ (A)} = (595 \pm 8) \text{ ns,}$$

and for class B

$$\Delta t \text{ (B)} = (668 \pm 4) \text{ ns.}$$

The time stability of LVD compared with that of the OPERA detector gives a time difference between the two classes

$$\Delta t \text{ (A} - \text{B)} = (-73 \pm 9) \text{ ns.}$$

This corresponds to a negative time shift for OPERA in the calendar period from August 2008 to December 2011 of the same order of the excess leading to a neutrino velocity higher than the speed of light as reported by OPERA [31].

Recent checks of the OPERA experimental apparatus showed evidence for equipment malfunctionings [32].

A first one is related to the oscillator used to produce the event time-stamps, while the second one is linked to the connection of the optical fiber bringing the GPS signal to the OPERA master clock.

This allows to conclude that the quantitative effect of this malfunctioning is the negative time shift, Δt (A – B), mentioned above.

This explains the previous OPERA finding [31] on the neutrino time of flight shorter by 60 ns over the speed of light.

The result of this joint analysis is the first quantitative measurement of the relative time stability between the two detectors and provides a check that is totally independent from the TOF measurements of CNGS neutrino events and from the effect presented in [32], pointing to the existence of a possible systematic effect in the OPERA neutrino velocity analysis.

If new experiments will be needed for the study of neutrino velocities they must be able to detect effects an order of magnitude smaller than the value of the OPERA systematic effect. This could be done [33] upgrading LVD with high TOF resolution [34].

## 8    ACKNOWLEDGMENTS

I would like to express my gratitude to the CERN Research Director, Sergio Bertolucci, and to the INFN President, Fernando Ferroni, for their wilful dedication to the solution of the problems which came up during the long period it took to finalise the results presented in this Report. These results required the collaboration of the members of the OPERA and LVD Groups, to whom I would also like to address my thanks, and in particular to Antonio Ereditato, Gabriella Sartorelli and the staff of the INFN-LNGS for his excellent work.

work by A. Petermann and A. Zichichi in which the renormalization group running of the couplings using supersymmetry was studied with the result that the convergence of the three couplings improved. This work was not published, but perhaps known to a few. The statement quoted is the first instance in which it was pointed out that supersymmetry might play an important role in the convergence of the gauge couplings. In fact, the convergence of three straight lines $(\alpha_1^{-1} \alpha_2^{-1} \alpha_3^{-1})$ with a change in slope is guaranteed by the Euclidean geometry, as long as the point where the slope changes is tuned appropriately. What is non trivial about the convergence of the couplings is that, with the initial conditions given by the LEP results, the change in the slope at $M_{SUSY} \sim 1$ TeV is not correct, as claimed by some authors not aware in 1991 of what was known in 1979 to A. Petermann and A. Zichichi.